# Generation of Secret Key for Physical Layer to Evaluate Channel Characteristics in Wireless Communications

B.U.V Prashanth[1,*], Y.Pandurangaiah[2]

[1]*Assistant Professor , Department of ECE,Vardhaman College of Engineering, Hyderabad,501218,AndhraPradesh,India*
[2]*Professor & Head , Department of ECE,Vardhaman College of Engineering, Hyderabad,501218,AndhraPradesh,India.*
*e-mail*: buv.prashanth@vardhaman.org, shiva.prashanth81@gmail.com

**Abstract**

This manuscript aims to generate a secret key for a PHY layer to evaluate the channel characteristics in wireless communications. An algorithmic approach is adopted for multimedia encryption to generate a secret key between two entities communicating with each other in a real time simulation environment such as MATLAB. Two classes of PHY key generation schemes are analyzed for designing the algorithm such as received-signal-strength-based and channel- phase-based protocols. We present a performance comparison of them in terms of key disagreement probability, key generation rate, key bit randomness, scalability, and implementation issues.

*Keywords*: PHY layer; eves dropping,;Secret key; Channel Characteristics;Wireless Communications;

## 1. Introduction

This paper aims to describe the process for encryption techniques for multimedia encryption. In this paper the multimedia image input is considered for encryption and decryption process. Traditional security schemes rely on public key infrastructures and cryptographic algorithms to manage secret keys. Recently, many physical-layer (PHY) based methods have been proposed as alternative solutions for key generation in wireless networks. These methods exploit the inherent randomness of the wireless fading channel to generate secret keys while providing information-theoretical security without intensive cryptographic computations. Due to the broadcast nature of wireless channels, wireless communication is vulnerable to eavesdropping, message modification, and node impersonation. In this manuscript we are calculating the scalability of the proposed method. Here we are encrypting the message using block matrix in two variables and then decrypting the message. Then scalability between input and output is calculated and displayed. In the next function the bit error rate is calculated and plotted on the graph. To this the random data is generated and the secret message is encrypted. In the decryption side the encrypted data is modulated with phase shifting keying and Additive White Gaussian Noise (AWGN) noise is added to the modulated data then after the Phase shift demodulation is applied to AWGN output. The bit error is calculated to the demodulated data and the final output is plotted on the graph.

## 2. Objective

Traditionally, security is an issue independent from physical layer in the 7 layers of the OSI Model and it relies heavily on the upper-layer operation. The symmetric data encryption/decryption algorithm has been widely used in networks. In this research manuscript we target at physical layer security to further enhance the wireless security [1]. Currently several attempts to PHY security include artificial noise technique which is designed to transmit a noisy signal to confuse eavesdropper but remains orthogonal to the channels between transmitter and receiver.

---

\* Corresponding author.

## 3. Security Analysis

In the following discussion the secret key derivation functions will be modeled in MATLAB v R 2012a starting with a random integer generation. Under this hypothesis, in the presence of a passive adversary run of the protocol can be easily seen to generate a uniformly distributed session key. Any principal A with access to an authentic copy of B can verify B's identity contained there in a straightforward manner [2]. However, A should verify B's identity during any protocol run to be assured that A will actually share a session key only with B. In this paper, we investigate two-party key agreement schemes based on asymmetric keys. In this paper we start our hypothesis by generating a prime number, followed by generating a public integer, secret integer for A and secret integer for B. In this we are modifying the Diffie Hellmann algorithm to further extent [5]. In this manuscript a block matrix is generated and is obtained from Bob. Next we generate a secret integer for encryption, and we formulate a secret matrix for encryption. Next we perform the operations such as read the input message, compress the input message, convert the compressed message into binary and obtain the block matrix of the message [3]. Now at this point the encryption process starts by considering by considering the number of block matrices. Now we obtain the normal matrix of the encrypted message. At this point the decryption routine starts by considering the beta value and alpha value which is considered as Ak1 obtained from Alice. Now the common secret value between Alice and Bob are to be considered in terms of a common key, if both common keys are matching then we display correct common key and provide the authentication towards the encryption side by entering a encryption key to view the message. For decryption we perform the inverse matrix operation [4]. The decryption is based on the following code snippet written in MATLAB R2012-a.Here we consider a block matrix as original matrix (orm1) as seen from below code snippet ,we perform the operations on the size and length as seen from the below code snippet.

```
blk2=blkmatx (orm1, 2);
l2=round (numel (orm1)/4);
p=0;q=0;z=1;
sz1=size (orm1);
for i=1:2:sz1(1)
for j=1:2:sz1(2)
len1r=i;
len2r=i+1;
len1c=j;
len2c=j+1;
if len2r-len1r==1&&len2c-len1c==1
for x=len1r:len2r
p=p+1;
for y=len1c:len2c
q=q+1;
if x<=length(orm1)&&y<=length(orm1)&&z<=l2
Q1(p,q,z)=orm1(x,y);
m4(p,q,z)=m2(x,y)
end
if q==2
q=0;
end
 end
 if p==2
p=0;
end
end
z=z+1;
 end
blk3(:,:,z-1)=Q1(:,:,z-1)*m4(:,:,z-1);
%decryption technique end
```

Further the scalability factor is described for applications where multiple parties are involved, a group key need to be established for securing the group communication [6].

*Generation of Secret Key For PHY Layer to Evaluate Channel Characteristics in Wireless Communications*## 4. Algorithm Design

The algorithm design is based on the steps as illustrated below
   a. Input  prime number
   b. Input  public integer (public key)
   c. Input secret integer of 'A'
   d. Input  secret integer of 'B'
   e. `Bk1=mod(g^beta1,N);%Bk1 obtained from Bob`
   f. Input secret integer for encryption
   g. secret matrix for encryption
   h. secret matrix with equal elements of message for encryption
   i. read the message
   j. compress the message
   k. Convert message in binary
   l. block matrix of message
   m. number of block matrices
   n. Encryption Routine starts
   o. normal matrix of encrypted message
   p. Decryption Routine starts
   q. enter the beta value(beta2)
   r. Ak1 obtained from Alice
   s. enter the common secret value
      The code snippet after entering the common secret value is as shown below.
```
 if cmk1==cmk3
disp('correct common key')
k2=input('enter encryption key(k2):')
if k2==c1/cmk2
Mk2=[1 1;k2 k2+1];
kM=inv(Mk2);%inverse matrix for decryption
m2=repmat(kM,50,50);
disp('correct encryption key')
else
disp('wrong encryption key')
end
```
   t. decryption technique
   u. After Decryption & before Round Off
   v. Disable the floating point numbers
   w. Find the Decrypted binary image
   x. Calculate the scalability
   y. Evaluate the characteristics of bit rate versus cumulative distribution function

## 5. Output Results

During the execution phase of this software design we have taken the input source as an image in .jpeg format for encryption purpose and the following lines describe the output results obtained from the MATLAB command prompt.
```
N =5392
Warning: This is an obsolete function and may be removed in the future. Please use
RANDI instead. To disable this warning, type warning ('off','comm: obsolete: randint').
alpha1 =8
beta1 = 9
k1 =21428
enter the beta value(beta2):9
beta2 =9
cmk2 =19032
```

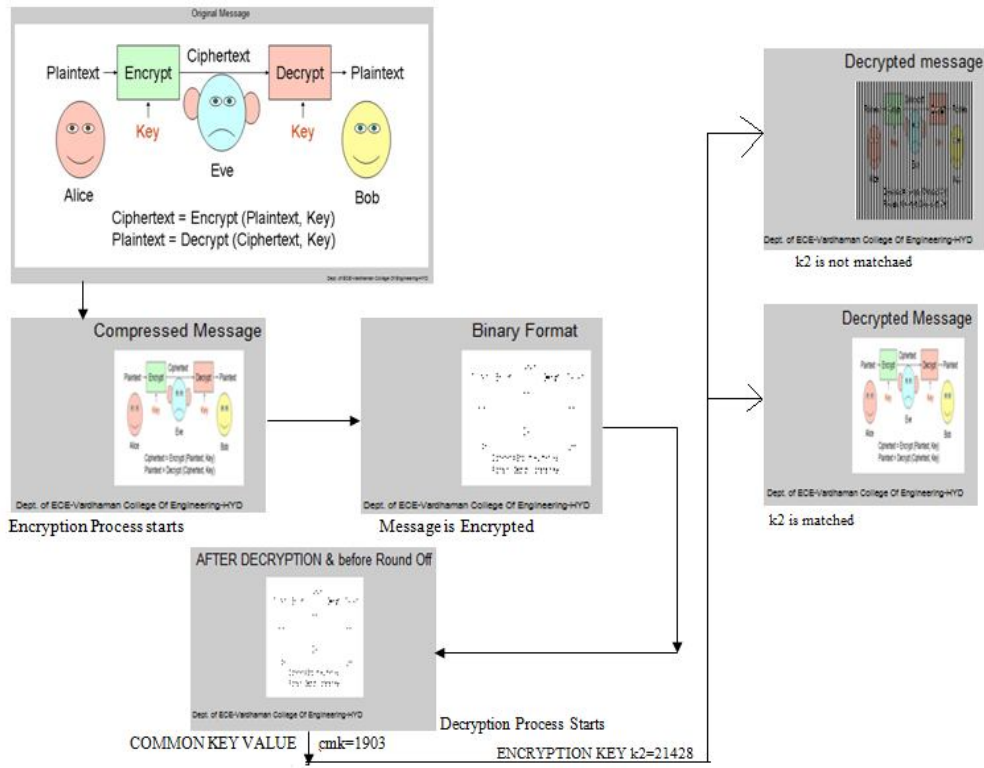

Figure 1 Flow Diagram of output results

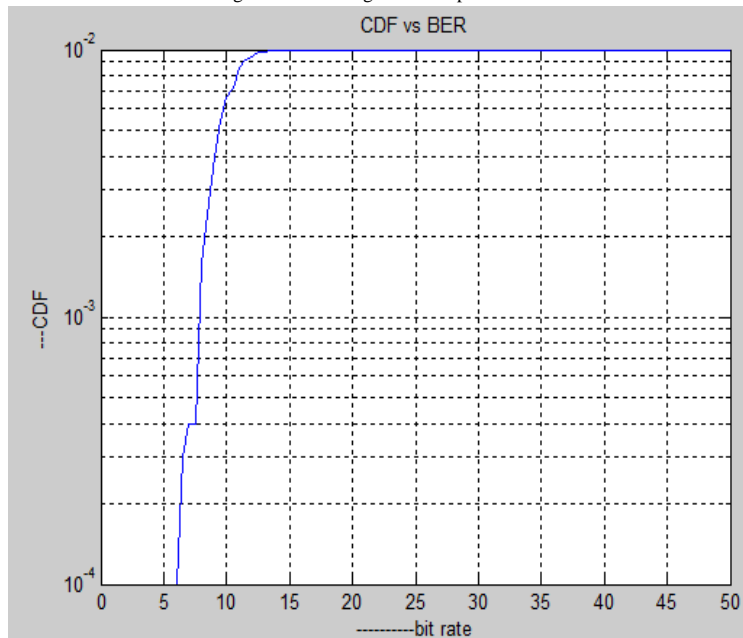

Figure 2 Plot between Cumulative Distribution Function and Bit rate

```
enter the common secret value:19032
cmk3 = 19032
correct common key
enter encryption key(k2):21428
k2 =21428
Correct encryption key
The scalability factor is 28902.5694
```

Here the N value is predefined along with predetermined values of alpha and beta. k1 is the secret key value which is having the dependencies based on alpha and beta values . Next we enter the common secret value; for the sake of simplicity for better understanding we have predefined common key value as 19032. As the common key values are matching we enter the secret value along with the verification of the correct common key, followed by entering the encryption key. Next we verify the correct encryption key. The following flow diagram in figure 1 illustrates the output results and also figure 2 illustrates the plot between Cumulative Distribution Function and Bit rate.

## 6. Conclusions

In this manuscript we have proven that the proposed protocol is secure in multimedia encryption technique. This protocol requires less computation cost for multimedia based cryptography applications. We have also shown that the proposed protocol is well suited for imbalanced wireless networks. A wireless communication system is accessed from the three aspects such as efficiency, reliability and security. A direction for research in security is explored for wireless channels from the point of view of providing features such as public key facility [7]. We have also determined the scalability factor as 28902.5694. The characteristics of cumulative distribution function versus the bit rate are found. In this paper the process for encryption techniques for multimedia encryption is analyzed. The multimedia image input is considered for encryption and decryption process [8]. The message is encrypted using block matrix in two variables and then the message is decrypted. Then scalability between input and output is calculated and displayed. In the next function the bit error rate is calculated and plotted on the graph. To this the random data is generated and the secret message is encrypted. In the decryption side the encrypted data is modulated with phase shifting keying and Additive White Gaussian Noise (AWGN) noise is added to the modulated data then after the Phase shift demodulation is applied to AWGN output. The bit error is calculated to the demodulated data and the final output is plotted on the graph. The output results obtained are found to be matching with the theoretical calculations. We obtained analytical results for this performance based algorithm and derived new insights, which led us to a conclusion that while asymmetric key exchange can generate an information-theoretic secret key under low noise, its efficiency degrades rapidly as a function of an adversary's (or external) signal interference power, thus limiting its resilience against active adversaries.

## 7. Future Scope

Toward a broader goal, we intend to develop wireless key exchange protocols that not only generate a strong key, but also generate good efficiency.